\documentclass[aip,pop,reprint]{revtex4-1}

\usepackage{geometry,amsmath,amssymb}             

\usepackage{graphicx,epstopdf,comment}

\newcommand{\pb}{$p$-$^{11}$B }
\newcommand{\pbend}{$p$-$^{11}$B}

\begin{document}

\title{Ignition threshold for non-Maxwellian plasmas}

\author{Michael J. Hay}
\email{hay@princeton.edu}
\affiliation{Department of Astrophysical Sciences, Princeton University, Princeton, New Jersey 08544}

\author{Nathaniel J. Fisch}
\affiliation{Department of Astrophysical Sciences, Princeton University, Princeton, New Jersey 08544}
\affiliation{Princeton Plasma Physics Laboratory, Princeton, New Jersey 08543}

\date{\today}
\begin{abstract}
An optically thin \pb plasma loses more energy to bremsstrahlung than it gains from fusion reactions, unless the ion temperature can be elevated above the electron temperature. In thermal plasmas, the temperature differences required are possible in small Coulomb logarithm regimes, characterized by high density and low temperature. The minimum Lawson criterion for thermal \pb plasmas and the minimum $\rho R$ required for ICF volume ignition are calculated. Ignition could be reached more easily if the fusion reactivity can be improved with nonthermal ion distributions. To establish an upper bound for this utility, we consider a monoenergetic beam with particle energy selected to maximize the beam-thermal reactivity. Channeling fusion alpha energy to maintain such a beam facilitates ignition at lower densities and $\rho R$, improves reactivity at constant pressure, and could be used to remove helium ash. The gains realized with a beam thus establish an upper bound for the reductions in ignition threshold that can be realized with any nonthermal distribution; these are evaluated for \pb and DT plasmas.
\end{abstract}

\maketitle

\section{Introduction}
Fusion reactions which release most of their energy in charged particles are desirable for power applications. In particular, if the number of energetic neutrons produced\cite{kernbichler} is small, the power plant can be designed with less expensive shielding and with fewer material constraints. Also, direct conversion\cite{post70} of charged particle energy could offer such a scheme much greater efficiency than a thermal cycle. The \pb reaction is ideal due to the low incidence of neutron-generating side reactions and the reactants' natural abundance. However, at typical densities, an optically thin \pb plasma loses more energy via bremsstrahlung radiation than it gains from fusion reactions, making sustained burn difficult.\cite{dawson,nevins98} Lacking sustained burn, or what is referred to as ignition, does not mean that fusion energy cannot be extracted, as in a wet wood burner. But the lack of ignition makes such a means of extracting energy technologically difficult and expensive.\cite{manheimer}

Previous efforts toward \pb ignition have focused on mitigating the bremsstrahlung loss channel. One approach is embedding large magnetic fields in the fusing plasma,  restricting the motion of unbound electrons and reducing the bremsstrahlung emission both directly, by modifying electron-electron scattering, and indirectly, by modifying electron-ion scattering. The Landau wavefunctions differ considerably from the isotropic, field-free case and the electron-electron scattering assumes a 1-D character. In the aggregate, the spectrally integrated emission is reduced by about 20\%.\cite{lauer83} Likewise, by increasing the Landau energy level spacing ($\hbar\Omega$), a hot ion mode ($T_i>T_e$) can be preserved due to the suppression of ion-electron energy transfer;\cite{miller87} such a mode is characterized by low bremsstrahlung emission at a given plasma pressure.

In a plasma with substantial electron degeneracy, bremsstrahlung is also reduced. The main effect appears to be a reduction of the electron stopping power,\cite{leon01,son04} enabling a large electron-ion temperature difference. For the same reason, degenerate electrons are preferable in beam-initiated detonations\cite{leon01, eliezer96, eliezer98} of compressed \pb fuel that rely on rapidly heating the ion species in a small locus.

Ignition using \pb is possible when the ion temperature is raised significantly above the electron temperature, a circumstance realized only in high-density thermal plasmas. (Though this does occur naturally in low temperature plasmas because of electron degeneracy.) $T_i\gg T_e$ is necessary, but not sufficient: $T_i$ itself must be on the order of several hundred keV due to the small \pb fusion cross section below that threshold. The resulting plasma is strongly collisional, but the robust fusion burn sustains ion temperatures above the electrons cooled by bremsstrahlung (note that $|T_i-T_e|\sim P_f/\nu_{ie}$ under these conditions).

Because high densities and temperatures are difficult to reach simultaneously, it is important to consider how either of these requirements might be relaxed. In equilibrium ignition-regime \pb plasmas, fusion alpha particles slow chiefly on the ions. Using waves to channel alpha energy\cite{fisch92,fisch94} to the lighter fusing species (protons) could improve the reactivity at fixed pressure, but, in the case of \pb as opposed to DT, the net gain is limited by the small fraction diverted to electrons in equilibrium.

Nonthermal distributions offer a plausible means to ease ignition requirements, either by improving the MHD stability of the bulk plasma,\cite{TA1,TA2} or by increasing the number of reacting particles near the cross section peak in velocity space, as treated here. This work considers the effect of such a non-Maxwellian feature in the ion distribution, namely a monoenergetic beam with particle energy maximizing the beam-thermal reactivity.

In order to discover the minimal ignition criteria, it is assumed that a fraction of the fusion power is efficiently channeled to maintain the beam against collisions with thermal particles. By way of comparison, these ignition conditions are calculated for both DT and \pb plasmas, with and without the presence of the monoenergetic beam, in both magnetic and inertial fusion configurations.

In all instances, power flows from a `hot' species (alphas, protons) to a cool one (protons, boron, respectively), and the extra power needed to maintain the nonthermal distributions is consistently included in the calculation. We explicitly include the effects of ion-ion and electron-ion drag in the energy balance.

However, we do neglect the power flow required to maintain the beam against velocity space diffusion. Consider a test particle born with energy slightly larger than that associated with the fusion cross section resonance. In both \pb and DT plasmas, this particle's fusion rate competes chiefly with slowing on light ions and electrons (boron excluded). Only after the particle slows out of the fusion resonance do the collisional drag and velocity space diffusion due to thermal ions become significant. Because the diffusion is especially small compared to drag at the beams' typical energies, we are justified in neglecting this term in the power balance (see Appendix~A).

Likewise, we do neglect kinetic effects such as collisionless streaming instabilities. A cold beam in a warm plasma will be subject to a complex array of instability mechanisms, but the two-stream instability between the fast monoenergetic beam ions and warm bulk species is prognostic.\cite{krall} However, in both DT and \pb ignition-regime plasmas, the bulk species are sufficiently cool with respect to the beam that nearly monoenergetic beams are resilient to this instability (see Appendix~A).\cite{friedwong}

Likewise, favorable assumptions and estimates are invoked to position the calculation as an upper bound on the feasibility of these fusion scenarios. For example, we explicitly assume that free energy in the alpha distribution can be transferred to the protons with perfect efficiency, whereas in practice rf power will likely be required to establish and maintain the mediating waves against collisions, etc. However, these beams do provide an upper bound on the gross gains (cf. net gains which would include the cost of maintaining a self-consistent distribution) realizable from nonthermal distributions; each additional particle in the beam adds the maximum amount of reactivity that could be gained from an extra particle at any energy.

In all cases, the relative locations of the thermal and beam reactivities' maxima determine the possible advantages of a nonthermal scheme. In particular, a successful beam contributes excess reactivity at constant pressure, so a plasma burning at a temperature close to the peak cross section energy (as in the case of DT) reaps limited benefit from nonthermal features. Although the igniting operating regime is difficult to access, the maxima of the \pb thermal and beam reactivities are germane to a nonthermal configuration which substantially lowers the ignition threshold. This work suggests the existence of nonthermal regimes where \pb ignition may be possible.

In section II, the model used to analyze the equilibrium state of a nonthermal plasma is detailed. Sections III and IV apply this model to predict ignition thresholds in magnetically- and inertially-confined plasmas. Section V considers the potential benefits of non-Maxwellian features in these plasmas.

\section{Thermal equilibration model}
In the case of \pb, the proton population is modeled as a thermal bulk plus a fast monoenergetic beam located near the \pb fusion cross section peak at 592 keV.\cite{nevins2000} A 0-D equilibrium model provides self-consistent species temperatures:
\begin{subequations}
\label{eq:system}
\begin{gather} 
\beta_b P_{\rm fus}= P_{\rm SD}\\[2mm]
\beta_e P_{\rm fus}+\nu_{ep}(T_p-T_e)\\ \nonumber
+\nu_{eB}(T_B-T_e)-P_{\rm brem}=0\\[2mm]
\beta_pP_{\rm fus}+\nu_{pe}(T_e-T_p)\\ \nonumber
+\nu_{pB}(T_B-T_p)=0\\[2mm]
 \beta_B P_{\rm fus}+\nu_{Bp}(T_p-T_B)\\ \nonumber 
 +\nu_{Be}(T_e-T_B)=0\\[2mm]
 G=G(\rho, \rho R;\, n_b, n_p, n_B;\, T_e, T_p, T_B)
\end{gather}
\end{subequations}
Eqs.~(1a-d) describe the flow of fusion charged particle power to each species; Eq.~(1e) is independent of the first four and determines the volume gain of an assembly with temperatures and densities characterized by (1a-d) once the scale of the system ($\rho$) is specified.

An effective frequency $\nu_{ij}=\nu_{ji}$ describes the drag of a thermal population $i$ on thermal population $j$.\cite{nrl} These coefficients are notably sensitive to the Coulomb logarithms characterizing the interacting species; the plasma electron density is an important parameter in establishing ignition criteria. The fusion power is the sum of a thermal reactivity\cite{nevins2000} and a beam-thermal reactivity,\cite{mikkelsen} viz. $P_{\rm fus}/W_f= n_B n_p\left< \sigma v\right>_{tt}+ n_B n_b\left< \sigma v\right>_{bt}$, where $W_f=8.7\,{\rm MeV}$ is the energy released in one reaction. The thermal bremsstrahlung emission $P_{\rm brem}$ is calculated as that from an optically thin medium, including relativistic corrections up to $(T_e/m_ec^2)^2$.\cite{rider}

The $\beta_i$ denote the fraction of the fusion power $P_{\rm fus}$ deposited in the $i^{\rm th}$ species, such that $\sum_i{\beta_i}=1$. In particular, $\beta_b$ is the fraction of the fusion power spent preserving the proton beam velocity distribution, such that $\beta_b=P_{\rm SD}/P_{\rm fus}$, with $P_{\rm SD}=\sum_i{n_b\nu_{bi}E_b}$ the amount of power dissipated by fast protons slowing down in the plasma (see Appendix~A). The $\nu_{bi}$'s are effective frequencies describing the fast beam particles' energy loss to the $i^{\rm th}$ species.\cite{nrl} The constant beam particle energy $E_b$ is displaced from the cross section peak to maximize the beam-thermal reactivity at a specified $T_B$. In order to limit the anticipated damping of waves used to construct the beam, $E_b$ was restricted to values at least twice as large as $T_B$.

In the large $T_e$ limit anticipated here, fusion alpha particles slow chiefly on the ions, and the amount of alpha power diverted to electrons is of order 10\%. In order to estimate the amount of alpha power absorbed by each species, consider a fast alpha particle born in a thermal \pb plasma with energy $E_i$. The amount of that particle's energy deposited in the protons is
\begin{gather}
\varphi _p=\frac{1}{E_i-E_f} \int\limits_{0}^{t_f}\! \nu_{\alpha p}E(t)\,dt
\end{gather}
where $E(t)$ is the instantaneous alpha particle kinetic energy. $\nu_{\alpha i}$ is the effective energy loss frequency for fast alpha particles slowing on the $i^{\rm th}$ species. $E_f=E(t_f)$ is the largest alpha energy at which energy flow from the particle to one species of field ions equals that from the field ions to the particle. As the typical alpha particle slows down from $\approx3\,{\rm MeV}$  in a \pb plasma with $T_p=T_B=T_i$, it reaches this dynamic equilibrium with the electrons first: although net energy exchange with the electrons has ended, the alpha particle continues to heat the proton and boron distributions at the $t_f$ defined this way. Thus, $E_f\approx1.5 T_e$, such that $\nu_{\alpha e}(t_f)$ vanishes. Changing the variable of integration to the particle energy, note 
\begin{gather}
\frac{dE}{dt}=-\sum_i{\nu_{\alpha i}E}
\end{gather}
and find
\begin{gather}
\varphi_p=\frac{1}{E_i-E_f}\int\limits_{E_f}^{E_i}\!  \frac{\nu_{\alpha p}}{\nu_{\alpha p} +\nu_{\alpha B}+\nu_{\alpha e}}\,dE
\end{gather}
$\varphi_B$ and $\varphi_e$ are defined analogously. This model satisfies $\varphi_B+\varphi_p+\varphi_e=1$; the distribution of fusion product energy is sensitive to the three species' temperatures as well as their relative concentration, determined by $\epsilon=n_A/n_B$, the number ratio of the heavy species `A' to the light species `B' (here $\epsilon=n_B/n_p$).

In the model problem of an equilibrium \pb reactor, where a fraction $\beta_b$ of the fusion power is devoted to counteracting the proton beam slowing down, only a fraction $\beta_p=(1-\beta_b)\varphi_p$ is available to the protons. This $\beta_b$ is extracted from alpha particles near their birth energies of several MeV. Because a temperature-dependent spectrum of birth energies peaking around 3.5 MeV is observed for the \pb reaction, it is assumed $E_i=3\,{\rm MeV}$ qualitatively predicts the energy flow to the thermal ions.\cite{stave} The model includes species absorption fractions calculated accordingly.

\section{Magnetic confinement}
The four equilibrium equations (1a-d) describe the time rate of change of the species temperatures and the amount of power spent preserving the monoenergetic proton beam. Because the fusion reaction rates and bremsstrahlung emission both depend on $n^2$, only the ratios of species densities ($n_b/n_p$ and $\epsilon=n_B/n_p$) are important. However, the temperature relaxation rates are sensitive to the Coulomb logarithm $\log \Lambda$, which in turn depends on the absolute electron density.

In order to explore the full space of igniting plasmas, consider a wide range of electron densities $(10^{10}<n_e<10^{30}\,{\rm cm}^{-3})$ and disregard the boron temperature equation (1d), instead taking $T_B$ as a specified parameter. The set (1a-c) can be solved for $\beta_b$, $T_e$, and $T_p$ once $n_e$, $T_B$, $n_b/n_p$, and $\epsilon$ are chosen. This specification of $T_B$ is tantamount to asserting an arbitrary ion energy confinement time.

With these assumptions, ignition thresholds can be defined for magnetically confined plasmas. Lawson criteria are presented for thermal \pb and DT plasmas, followed by nonthermal \pb and DT plasmas. These latter calculations provide the upper bounds on the utility of maintaining a nonthermal distribution, abiding by the assumptions laid out in the Introduction. In the thermal cases, the numerical model predictions are compared to analytic calculations.

\subsection{Lawson criterion}
A common metric for the performance of fusion power systems is the criterion first obtained by Lawson.\cite{lawson} In steady state, a plasma ignites if  the $n\tau$ product of number density and confinement time exceeds a specific value depending on the fusion reaction under consideration. The Lawson criterion is the statement of this minimum value, determined simply by $P_{\rm fus}>P_{\rm loss}$. Taking $P_{\rm fus}=W_f n_A n_B\left<\sigma v\right>$, with $W_f$ the energy released in charged particles, and defining $\tau=W/(-dW/dt)$ where $W$ is the thermal energy content of the plasma and $-dW/dt=P_{\rm brem}=P_{\rm loss}$,\cite{freidberg}
\begin{gather}
\label{lawson}
n_e \tau \geq \frac{T(Z_B+Z_A\epsilon)(1+Z_B+\epsilon(1+Z_A))}{W_f \left<\sigma v\right>\epsilon}
\end{gather}
where $\epsilon= n_A/n_B$, $n_e=Z_An_A+Z_Bn_B$, and $T$ is the temperature common to all species. In general, the species temperatures differ, rendering the minimization of the right hand side nontrivial. It is however instructive to search numerically for those operating conditions which afford the least stringent Lawson criterion. The resulting numerical criterion is necessary but not sufficient for ignition. In pressure-limited systems, the triple product $nT\tau$ is a superior metric because it is proportional $T^2/\left<\sigma v\right>$, in turn inversely proportional to the achievable fusion power $W_f p^2\left<\sigma v\right> /T^2$.\cite{shultis} Thus the threshold igniting state has minimum $nT\tau$ and maximum power.

\subsection{Minimum Lawson criterion for DT ignition}
Assuming $T_e=T_i=T$, one can make a simple estimate of the threshold Lawson criterion for DT ignition. Because both deuterium and tritium carry the same number of electrons, $P_{\rm brem}$ does not depend directly on the number ratio $\epsilon$. The fusion power, however, is maximized for $\epsilon=n_T/n_D=1$, so the extremal ignition case will be located near that operating point. Using Eq.~\eqref{lawson}, we can form the triple product criterion for DT:
\begin{gather}
nT\tau\geq \frac{2T^2}{W_f \left<\sigma v\right>}\frac{(1+\epsilon)^2}{\epsilon}= f(\epsilon)g(T).
\end{gather}
In other words, the minimum triple product is a separable product of functions of $\epsilon$ and $T$. Considering minimization with respect to $\epsilon$, we have
\begin{gather}
{\rm min}(nT\tau)\propto2+\epsilon+\frac{1}{\epsilon},
\end{gather}
which has its minimum at $\epsilon =1$. The criterion is also proportional to $T^2/\!\left<\sigma v\right>$, a function of temperature only. The numerical optimum temperature should be located close to the criterion's minimum at $T=13.5\,{\rm keV}$. The estimated lower bound is $1.9\times 10^{21}\,{\rm keV}\!\cdot{\rm s}/{\rm m}^3$.

Alternately, the consistent set given by Eqs.~(1a-d) can be solved at various $T_i$ to determine the threshold (minimum $nT\tau$) igniting configuration. A numerical search found the extremal point at a mass-weighted ion temperature $T_{\rm ion}=8.7\,{\rm keV}$ and $\epsilon=0.97$. The corresponding minimum value of $nT\tau$ is $2.6\times 10^{21}\,{\rm keV}\!\cdot{\rm s}/{\rm m}^3$, in close agreement with the estimated value. A slight excess of deuterons is explained by their more efficient use of thermal energy. At fixed energy, the lighter particles have larger velocities, and for a thermal 9 keV DT plasma, larger cross sections ($\sigma_f$ is monotone increasing up to $E_{CM}=64\,{\rm keV}$). Balancing this effect are the fusion rate penalty with $\epsilon\neq1$,
\begin{gather}
P_{\rm fus}\propto\frac{\epsilon}{1+\epsilon},
\end{gather}
and deuterons' larger drag losses to the electrons; the ratio of temperature equilibration rates is $\nu_{De}/\nu_{Te}\approx1.5$.

\subsection{Minimum Lawson criterion for \pb ignition}\label{pbthermal}
Following the approach of the previous section, the triple product criterion has the density scaling
\begin{gather}
{\rm min}(nT\tau)\propto8+\frac{1}{\epsilon}+15\epsilon,
\end{gather}
which takes its minimum value at $\epsilon=n_B/n_p=1/\sqrt{15}\approx0.26$. Likewise, the quantity $T^2/\!\left<\sigma v\right>$ is minimized for $T=138\,{\rm keV}$. This inconsistent optimization, regarding $\epsilon$ and $T$ as totally independent quantities, returns an estimated lower bound of $5.1\times 10^{23}\,{\rm keV}\!\cdot{\rm s}/{\rm m}^3$. However, a \pb plasma cannot ignite with $T_e=T_i$. (That is, unless a very substantial number of x-rays can be reflected off the walls and reabsorbed by the plasma. Here, we assume that for all practical purposes, this cannot be done. We likewise assume that the plasma is optically thin to bremsstrahlung.) Per the analytic model, $P_{\rm fus}/P_{\rm brem}$ has a maximum value of $0.44$ at $T_e=T_i=204\,{\rm keV}$ and $\epsilon=0.11$.

A numerical search allowing for species-dependent temperatures found the extremal ignition point at a mass-weighted ion temperature $T_{\rm ion}=193\,{\rm keV}$ and $\epsilon=0.26$. The corresponding minimum value of $nT\tau$ is $2.2\times 10^{23}\,{\rm keV}\!\cdot{\rm s}/{\rm m}^3$, in close agreement with the estimated value. The numerical optimum is cooler and significantly more boron-rich than the na\"ive minimization of Eq.~\eqref{lawson} would suggest. The extremal ignition state offsets a larger boron concentration (more fusions and bremsstrahlung) with a cooler $T_e=59\,{\rm keV}$.

\subsection{Nonthermal gains: DT}

Suppose there were some way, say by alpha channeling, to support a monoenergetic deuterium beam. In both the DT and \pb plasmas, the lighter species is chosen for the beam because of the lower energy investment required to achieve the high center of mass energy needed to access the fusion cross section peak. Operating points with $\lambda\doteq\log\Lambda<3$ were discarded because the coupling coefficients $\nu_{ij}$ are accurate only to first order in $\lambda^{-1}$. Points with $\beta_b>1$, signaling the need for injected power to maintain the beam, were discarded likewise.

The maximum reactivity subject to a pressure constraint is a good metric for comparing the effects of nonthermal distributions in fusion plasmas because the pressure a reactor can confine is limited by magnet strength. In the case of ITER, this figures to be about 10 bar, which we adopt as a standard value in order to compare DT and \pb MFE schemes. Nonthermal features which improve the fusion reactivity at a fixed or reduced pressure are therefore desirable. In general, the plasma pressure includes contributions from any beam as well as the thermal populations and any alpha particles slowing down on thermal particles. Because the alpha particles whose energy is channeled to maintain the beam (fraction $\beta_b$) are `lost' on a fast, collisionless timescale, their pressure can be assumed to be negligible. The total alpha pressure is estimated as
\begin{gather}
(1-\beta_b)n_\alpha \frac{W_f}{N_\alpha}=(1-\beta_b)\frac{P_{\rm fus}}{N_\alpha \nu_{\rm SD}}
\end{gather}
where $n_\alpha$ is the number density of alpha particles, $N_\alpha$ is the number of alpha particles spawned by a single reaction, and $\nu_{\rm SD}$ is the slowing down frequency on thermal particles. In practice, wave-mediated diffusion and device confinement could supersede Coulomb collisions with thermals as the salient processes limiting the average alpha lifetime. The reactivity includes contributions from both thermal-thermal reactions and beam-thermal reactions.

The ion number ratio which maximizes the reactivity of a pressure-limited, constant-temperature system (in the absence of a beam) is (see Appendix~B)
\begin{gather}
\epsilon_0=n_A/n_B=(1+Z_B)/(1+Z_A).
\end{gather}
The temperature $T_0$ at the reactivity maximum satisfies
\begin{gather}
\frac{d}{dT}\left<\sigma v\right>\!(T)\bigg|_{T=T_0}=2\frac{\left<\sigma v\right>\!(T_0)}{T_0}
\end{gather}
Note that both of these conditions are independent of the limiting pressure. In the case of a constant-temperature DT plasma, the optimal reactivity is found at $T=13.5\,{\rm keV}$ and $\epsilon=1$. In the limit of small beam fraction $\varphi\to0$, the numerical model locates the reactivity optimum under a 10-bar pressure constraint at $T_i=15.3\,{\rm keV}$ and $\epsilon=0.89$.

The introduction of a fast deuterium beam at the cross-section peak can improve the fusion reactivity at constant pressure. Fig.~1 indicates the possible gains up to a beam fraction of 1/2. The system pressure is held fixed at 10 bar. 

As the beam fraction increases, the reactivity-optimizing temperature decreases and the ion ratio tilts toward the heavier target species. Both of these shifts tend to increase the beam reactivity at the expense of thermal reactivity. Such an optimized beam reduces the necessary thermal energy content of an igniting system because fewer fast particles are off-resonance; at constant pressure, a larger reactivity is possible. Likewise, a fast beam reduces the density and pressure required for an igniting plasma, substantially easing the minimum ignition conditions. 
\begin{figure}
\includegraphics[width=75mm]{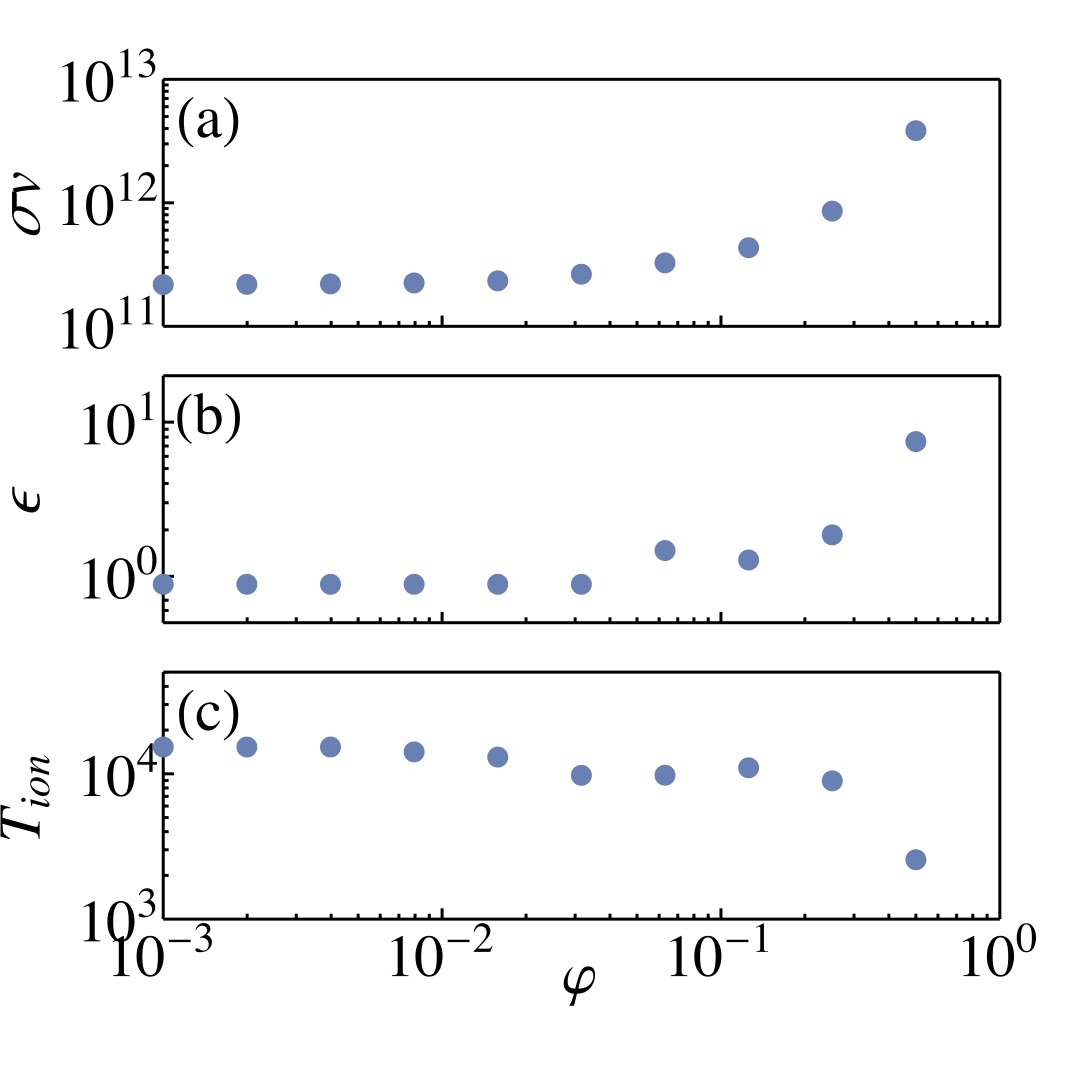}
\caption{Optimized DT operating points with increasing nonthermal features in the light ion distribution function. The deuterium beam fraction $\varphi$ varies from $10^{-3}$ to 0.5; the optima at larger beam fractions are characterized by greater reactivity (a) $[{\rm cm}^{3}/{\rm s}]$, increased target ion concentration (b), and lower ion temperatures (c) [eV].}
\end{figure}

\subsection{Nonthermal gains: \pb}
In the beam-free, Maxwellian case, subject to a pressure constraint, the \pb reactivity is maximized at $T=138\,{\rm keV}$ and $\epsilon=1/3$. In the limit of small beam fraction, the reactivity maximum (under a 10-bar pressure constraint) occurs at $T_i=137\,{\rm keV}$ and $\epsilon=0.20$. The discrepancy in $\epsilon$ is due to the significant bremsstrahlung emission present in the full model at higher values of $\epsilon$. The possible gains from the addition of a fast proton beam are indicated in Fig.~2, up to a beam fraction of 0.1. 

The energy cost of maintaining a given beam fraction against collisions with thermals is comparatively greater in \pb than DT, due primarily the abundance of electrons. Above about $\varphi=0.2$, a proton beam cannot be maintained without injected power. In fact, above beam fractions of $0.02$, these beams were only useful at higher temperatures, where the slowing down of beam protons is reduced, so the beams can be maintained at lower cost. Likewise, in the absence of a pressure constraint, fast beams can be used quite profitably. At sufficiently high temperature and density, the power required to counteract beam drag is a small fraction of the total thermal power.

\begin{figure}
\includegraphics[width=75mm]{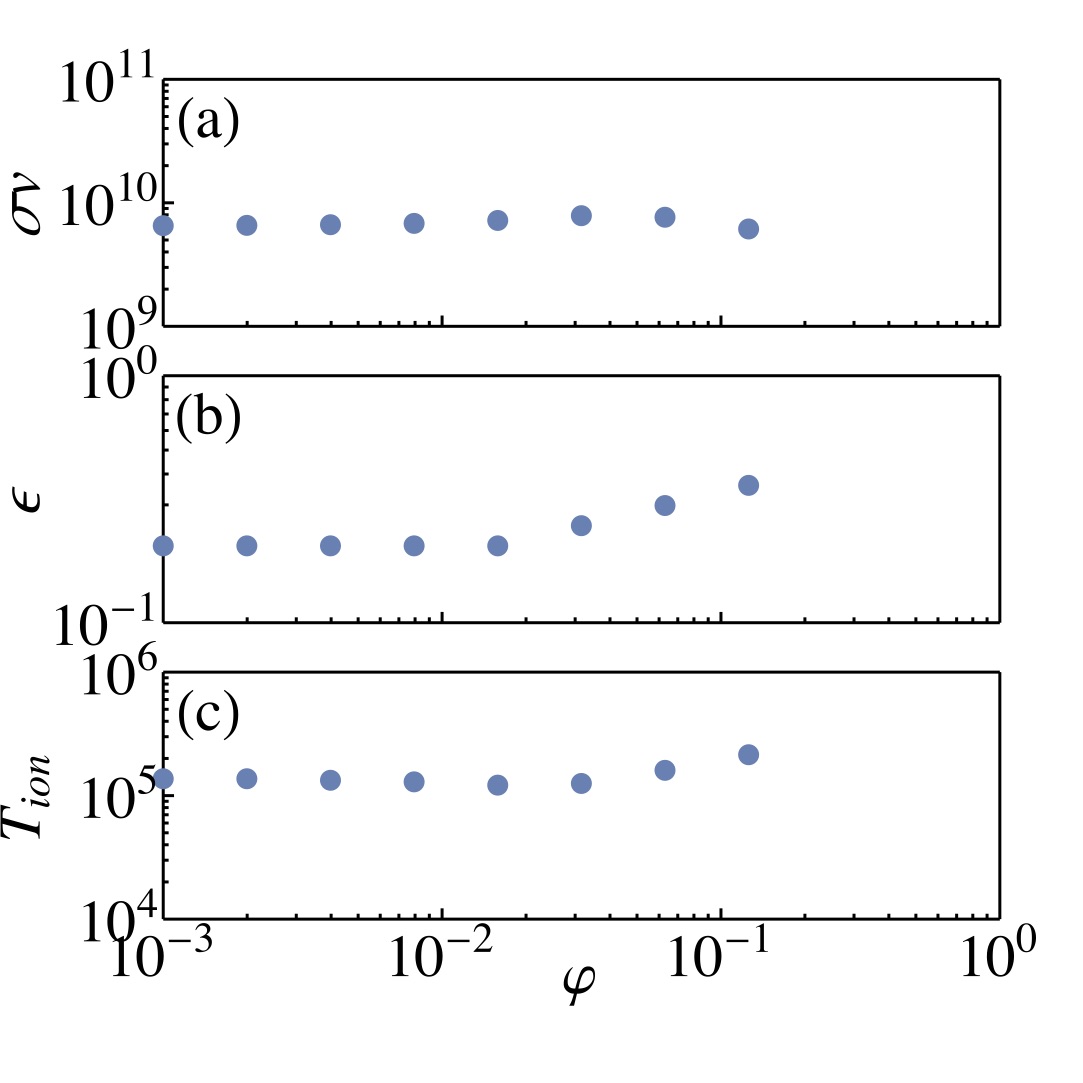}
\caption{Optimized \pb operating points with increasing nonthermal features in the light ion distribution function. The proton beam fraction $\varphi$ varies from $10^{-3}$ to $10^{-1}$; the optima at larger beam fractions are characterized by greater reactivity (a) $[{\rm cm}^{3}/{\rm s}]$, increased target ion concentration (b), and higher ion temperatures (c) [eV].}
\end{figure}

The removal of fast alpha particles in the channeling process is a significant boon to \pb ignition prospects. In a pressure-limited device, the enormous number of fusion alpha particles produced poisons the reaction by contributing pressure without also providing reactivity. Even a marginal increase in the reactant pressure, as would be the case with a small beam fraction of fast particles, could be helpful because the fusion power scales as $p^2$.

Consider the timescale defined by $\mathcal{T}=p_{th}/P_\alpha$, where $p_{th}$ is the total pressure of all thermal particles in the plasma and $P_\alpha= \epsilon_\alpha \dot{n}_\alpha$ is the instantaneous fusion power released in alpha particles. If the pressure $p_{th}$ varies slowly on the timescale $\mathcal{T}$, for example in the case of continuous refueling, $\mathcal{T}$ is a reasonable estimate of the time a plasma can burn in quasi-steady state conditions (for which alpha poisoning is insignificant). The ratio $\mathcal{T}_{\rm DT}/\mathcal{T}_{pB}$ is of interest. Assuming equimolar DT with $T_D=T_T=T_i$,
\begin{align}
\mathcal{T}_{\rm DT}&=\frac{n_D T_D+n_T T_T+n_e T_e}{\epsilon_\alpha n_D n_T \left<\sigma v\right>}\\ 
&\approx \frac{4(T_i+T_e)}{\epsilon_\alpha n_e\left<\sigma v\right>}
\end{align}
Supplying projected ITER plasma parameters, $\mathcal{T}_{\rm DT}\approx 1\,{\rm s}$, which indicates that controlling alpha pressure will be important during pulses expected to last several minutes. In comparison,

\begin{align}
\mathcal{T}_{pB}&=\frac{n_p T_p+n_B T_B+n_e T_e}{3\epsilon_\alpha n_p n_B \left<\sigma v\right>}\\ 
&= (1+5\epsilon)\frac{(T_p+T_B+(1+5\epsilon)T_e)}{3\epsilon_\alpha n_e\left<\sigma v\right>},
\end{align}
where $3\epsilon_\alpha=8.7\,{\rm MeV}$. Note that in the case of the \pb reaction, $P_\alpha=P_{\rm fus}$. Supplying the parameters minimizing the Lawson criterion for a thermal \pb plasma (see section~\ref{pbthermal}), $\mathcal{T}_{pB}\approx 10^{-16}\,{\rm s}$. Clearly an active means of removing alpha pressure (on a collisionless timescale) is crucial for any plausible \pb reactor.

Although the potential gains of a fast beam in \pb plasmas (due to the high reactivity of a beam near the thermal bulk) are limited by drag on the electron densities required, alpha channeling may yet prove an invaluable means of controlling the alpha poisoning effect.
\section{Inertial confinement}
Here we describe the volume ignition scheme\cite{atzeni} of inertial fusion and discuss how DT and \pb plasma parameters might be optimized to lower the ignition threshold. Although conservative in its predictions (cf. practical ICF schemes), the volume model is useful because of its simple structure (nearly neglecting hydrodynamic motion), highlighting the effect of a fast beam on ignition conditions. In particular, we will determine to what extent fast beams could reduce the assembly energies and $\rho R$ of DT and \pb volume targets.

The gain equation (1e) is independent of the equilibrium equations (1a-d). To solve the gain equation, however, the absolute density $\rho$ and scale ($\rho R$, or equivalently $E_a$) of the system must be specified. Together with the temperatures and beam density supplied by Eqs.~(1a-d), these are sufficient to evaluate the volume gain (the yield of a homogeneous spherical assembly burning in a sound time divided by its initial thermal energy). 

\subsection{Volume ignition}
The volume ignition scheme \cite{atzeni} imagines a spherical target that has been prepared in a completely homogeneous state at the time of ignition. This provides a conservative estimate of the obtainable gain because the entire fuel must be heated; more tractable ICF schemes rely on the heating of only a small portion of the burning mass, reducing the total thermal energy of the assembly, $E_a$. In practice, the driver energy required to assemble the target is greater than $E_a$ due to backscatter, x-ray conversion losses, solid angle effects, rocket efficiency, etc.

The gain associated with a volume-ignited target can be expressed simply as (cf. Eq.~(1e))
\begin{align}
G=\frac{E\, \phi}{\frac{3}{2}\Gamma\, T_{\rm eff}}\label{eq:volumegain}
\end{align}
where $E$ is the fusion energy released per unit mass ($3.39\cdot10^{11}\,\rm{J/g}$ for DT, $7\cdot10^{10}\,\rm{J/g}$ for \pb) and $\phi$ is the fraction of the target mass that burns before hydrodynamic disassembly. $\Gamma=k_B/m$ is the plasma's specific gas constant ($m$ being the mean mass of the constituent particles) and $T_{\rm eff}$ is the number-weighted temperature. $\frac{3}{2}\Gamma\, T_{\rm eff}=E_a/M_f$, where $M_f$ is the total target mass. It is assumed that classical statistics suffice to describe each of the target's constituent species. 

In order to operate in a high-gain regime, a typical ICF target will burn a significant fraction of its fuel.\cite{lindl95} It is therefore necessary to integrate the fusion rate equation $\dot{n}\sim n^2$ over the confinement time to estimate the fraction of fuel consumed in fusion reactions. The result for a mixture of two species with initial number densities $n_{A0}$ and $n_{B0}$, such that $\epsilon\doteq n_{A0}/n_{B0}$, is (see Appendix C)

\begin{align}
\phi=\frac{2}{\epsilon+1}\left[1-\frac{\epsilon-1}{\epsilon\exp\!\left( 2\frac{\epsilon-1}{\epsilon+1}\frac{\rho R}{H_B}\right) -1}\right]
\end{align}
where $H_B=6 c_s m/\!\left<\sigma v\right>$, $c_s=(T_e/m)^{1/2}$ is the ion sound speed, and $m=(\epsilon m_A+m_B)/(\epsilon+1)$. $\rho=m_A n_{A0}+m_B n_{B0}$ is the initial mass density. In the limit $\epsilon\to1$, the familiar $\rho R$ formula for an equimolar target is recovered: 

\begin{align}
\phi\to\frac{\rho R}{\rho R+H_B}
\end{align}
In the limit $\rho R\to\infty$, burnup is limited to $2/(\epsilon+1)$.

\subsection{Optimization procedure}

Along with $\epsilon$, the initial ratio of boron to hydrogen nuclei, the proton beam fraction $n_b/n_p$ is regarded as a parameter. With $\epsilon$ and $n_b/n_p$ specified, the system of Eqs.~\eqref{eq:system} can be solved for $T_e$, $T_B$, $T_p$, and $n_b$. These state variables are used to evaluate the volume gain for the configuration, which is then a function of $\rho R$ only (through the burn fraction, $\phi$). A large sampling space was considered to locate the minimum assembly energy $E_a$ necessary for a volume gain of unity (Eq.~\eqref{eq:volumegain}). $E_a$ can be related directly to the equilibrium conditions determined by Eqs.~(1a-d):
\begin{gather}
E_a\doteq\frac{3}{2}NT_{\rm eff}=\frac{3}{2}\sum{N_s T_s}\\
=\frac{3}{2}\frac{M_f [T_e+T_p+\epsilon(5T_e+T_B)]}{m_e+ m_p+\epsilon(5m_e+m_B)}
\end{gather}
in the case of \pb with $n_{B0}/n_{p0}=\epsilon$ and $n_b/n_p=0$ (no beam). $M_f$ is the mass of the entire target; Boltzmann's constant $k_B$ has been suppressed. Using
\begin{align}
M_f=\frac{4\pi(\rho R)^3}{3\rho^2}
\end{align}
the burnup $\phi=\phi(\rho R, T_e, T_p,T_B)$ and therefore the volume gain are entirely specified by $E_a$ once the assembled density $\rho$ has been given (in addition to $\epsilon$ and $\beta_b$). To wit,
\begin{align}
\rho R = \left(\frac{\rho^2 E_a}{2\pi}\frac{m_e+m_p+\epsilon (5m_e+m_B)}{T_e+T_p+\epsilon(5T_e+T_B)}\right)^{1/3}
\end{align}
The addition of a fast proton or deuteron beam is reflected by 
\begin{align}
E_a\to \frac{3}{2}\left(n_b\epsilon_b+\sum{N_s T_s}\right)
\end{align}
Thus, once equations (1a-d) have been solved consistently for the species temperatures with specified $T_B$, $\epsilon$, beam fraction, and density, the gain equation (1e) can be solved for the minimum assembly energy resulting in a volume gain of unity. Using the above definitions, it is possible to express this criterion in terms of the assembly $\rho R$. 

\subsection{DT ignition criterion}
Assuming a constant temperature $T$ for all species, it is straightforward to estimate a beam-free best case from the gain equation (1e). At $\rho=10^3$, the result is a minimum igniting $\rho R$ of $0.022\,{\rm g/cm}^2$ at $T=18\,{\rm keV}$ and $\epsilon=n_T/n_D=1.10$. This is equivalent to an assembly energy of 0.09 J.

A numerical search allowing for distinct, consistent species temperatures determined by the system (1a-e) located a minimum $\rho R$ of $0.07\,{\rm g/cm}^2$ at a mass-weighted ion temperature of $7.1\,{\rm keV}$ and $\epsilon=1.13$. This is equivalent to an assembly energy of 1.1 J. In the full model, slowing of fusion alpha particles on the thermal electrons creates some separation in electron and ion temperatures ($T_e\approx10\,{\rm keV}$ here), rendering the assumptions of the original estimate inaccurate.

In the presence of a fast deuteron beam, the ignition criteria are further relaxed. The typical minimal state features a fast beam colliding with a thermal plasma substantially cooler than the Maxwellian optimum. In DT, the potential utility of the beam is limited by the fact that the resonant beam energy is an order of magnitude greater than the optimal Maxwellian plasma temperature whereas the reactivity gain is limited to a factor of about 3.5. As Fig.~3 indicates, a fast deuteron beam nonetheless reduces the $\rho R$ required for ignition by a factor of 14, corresponding to a factor of $10^4$ reduction in the total thermal energy of an igniting assembly. Above a beam fraction of about 4\%, the shift towards large tritium concentrations (Fig.~3b) and low ion temperatures (Fig.~3c) is pronounced.

\begin{figure}
\includegraphics[width=75mm]{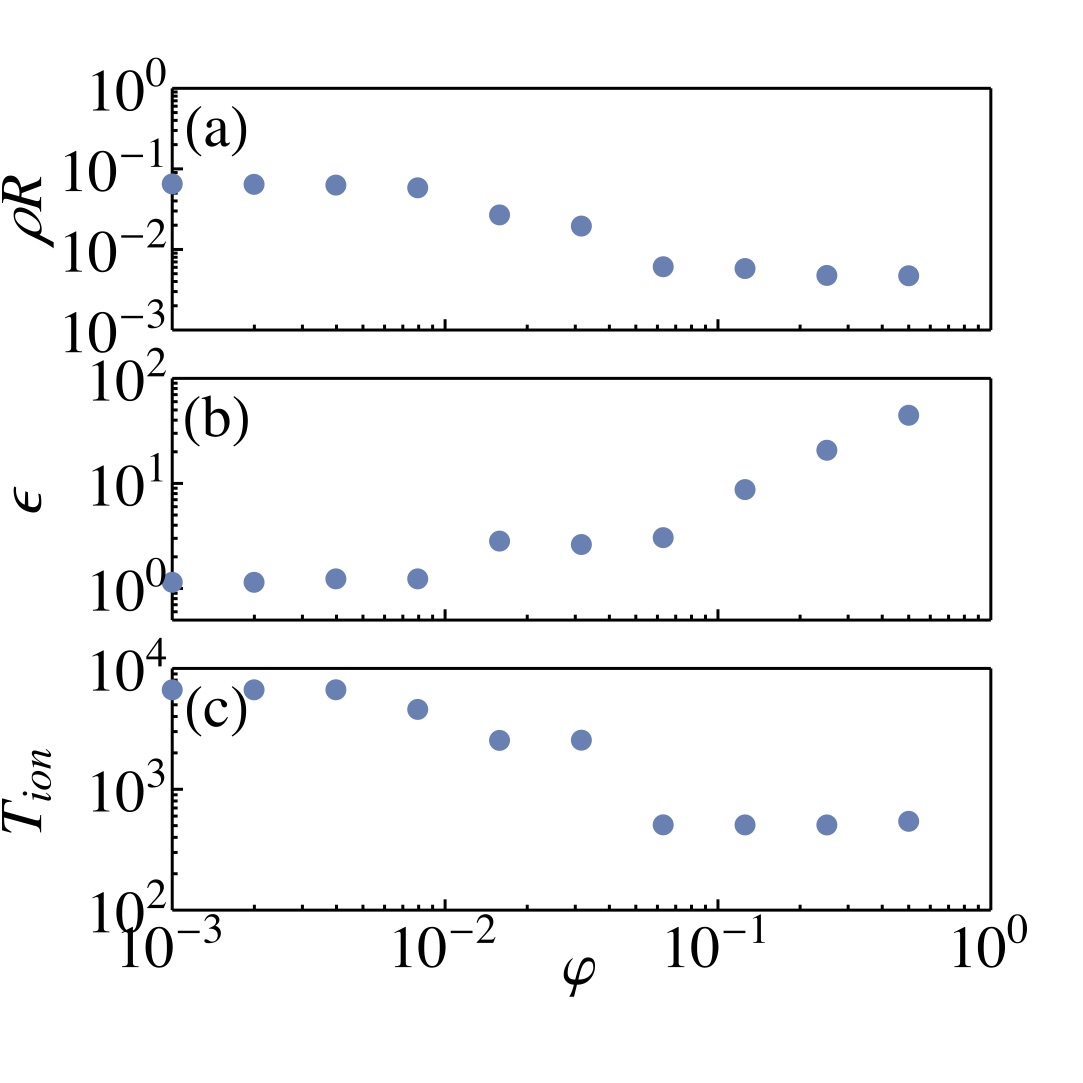}
\caption{Optimized DT operating points with increasing nonthermal features in the light ion distribution function. The deuteron beam fraction $\varphi$ varies from $10^{-3}$ to $0.5$; the optima at larger beam fractions are characterized by lower minimum igniting $\rho R$ (a) $[{\rm g\cdot cm}^{-2}]$, increased target ion concentration (b), and lower ion temperatures (c) [eV].}
\end{figure}

\subsection{\pb ignition criterion}
Assuming a constant temperature $T$ for all species, it is straightforward to estimate a beam-free best case from the gain equation (1e). At $\rho=10^3$, the result is a minimum igniting $\rho R$ of $9.95\,{\rm g/cm}^2$ at $T=153\,{\rm keV}$ and $\epsilon=n_B/n_p=1.30$. This is equivalent to an assembly energy of 59 MJ.

A numerical search allowing for distinct, consistent species temperatures determined by the system (1a-d) located a minimum $\rho R$ of $2.1\,{\rm g/cm}^2$ at a mass-weighted ion temperature of $196\,{\rm keV}$ and $\epsilon=1.8$. This is equivalent to an assembly energy of 0.28 MJ, showing marked improvement over the na\"ive estimate. At the high densities characteristic of ICF, the Coulomb logarithm is small enough that a large temperature difference can be sustained between electrons and ions. In the particular case of \pb, the substantial bremsstrahlung emission keeps the electron temperature low. The resulting lower thermal content of the fusing plasma improves the volume gain. 

In the presence of a fast proton beam, the ignition criterion is relaxed. In contrast to DT, the resonant beam energy (c. 592 keV) is only a factor of 2-4 greater than the optimal Maxwellian plasma temperature, whereas the reactivity gain is superior, about a factor of 6. Fig.~4 suggests that in \pb, the net effect of a beam  is nevertheless less pronounced cf. DT. The minimum igniting $\rho R$ is reduced by nearly a factor of 3 (Fig.~4a), corresponding to a factor of 30 reduction in assembly energy. The optima eschew contributions from thermal reactions above a beam fraction of about 7\%, as demonstrated by the shift to larger $\epsilon$ (Fig.~4b) and low ion temperature (Fig.~4c).

\begin{figure}
\includegraphics[width=75mm]{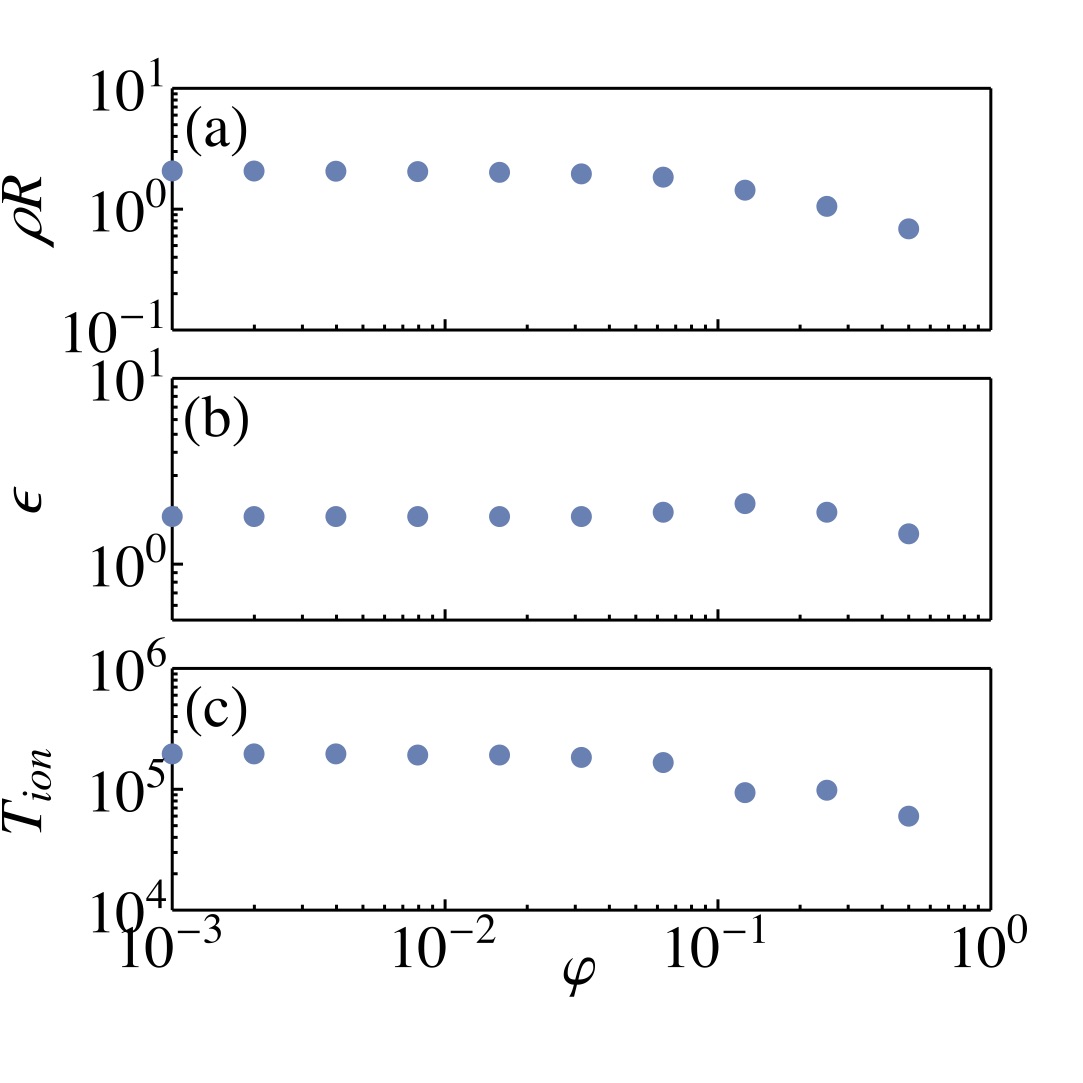}
\caption{Optimized \pb operating points with increasing nonthermal features in the proton distribution function. The proton beam fraction $\varphi$ varies from $10^{-3}$ to $0.5$; the optima at larger beam fractions are characterized by lower minimum igniting $\rho R$ (a) $[{\rm g\cdot cm}^{-2}]$, roughly equal target ion concentration (b), and lower ion temperatures (c) [eV].}
\end{figure}

\section{Discussion}
This work seeks to address, in broad strokes, the utility of non-Maxwellian features in both DT and \pb plasmas under magnetic and inertial confinement. Regardless of the scheme chosen, the relation between the thermal reactivity and resonant features in the fusion cross section is crucial in establishing this utility.

A natural criterion in both confinement schemes is the amount of pressure or thermal energy required for ignition because this quantity scales directly with the facility cost (magnet strength in MCF, driver energy in ICF). If a beam is to reduce the pressure requirement, it should provide excess reactivity without a disproportionate contribution to the system pressure. In particular, the beam-thermal reactivity ratio and the ratio of beam energy to the bulk temperature should be compared. Heuristically, the parameter
\begin{gather}
R=\frac{\left<\sigma v\right>_{\rm beam}/\!\left<\sigma v\right>}{\epsilon_{\rm beam}/T}
\end{gather}
is indicative of the beam utility for specified plasma conditions.

In the case of DT, the fusion resonance is located far in the tail of most igniting plasmas (true `thermonuclear' fusion). Because bremsstrahlung losses are minimal, the plasma energy content required for ignition is small (with respect to the resonance), but the distance of the resonance is a drawback when energy is supplied directly to heat particles at such large energies, as in the case of a resonant beam. Likewise, the broad DT resonance limits the achievable reactivity gain; the value of $R$ in igniting thermal DT plasmas is about 1/2, indicating that channeling fusion power to a beam is not generally useful. Fig.~5 traces $R$ for constant-temperature DT and \pb plasmas.
\begin{figure}
\includegraphics[width=65mm]{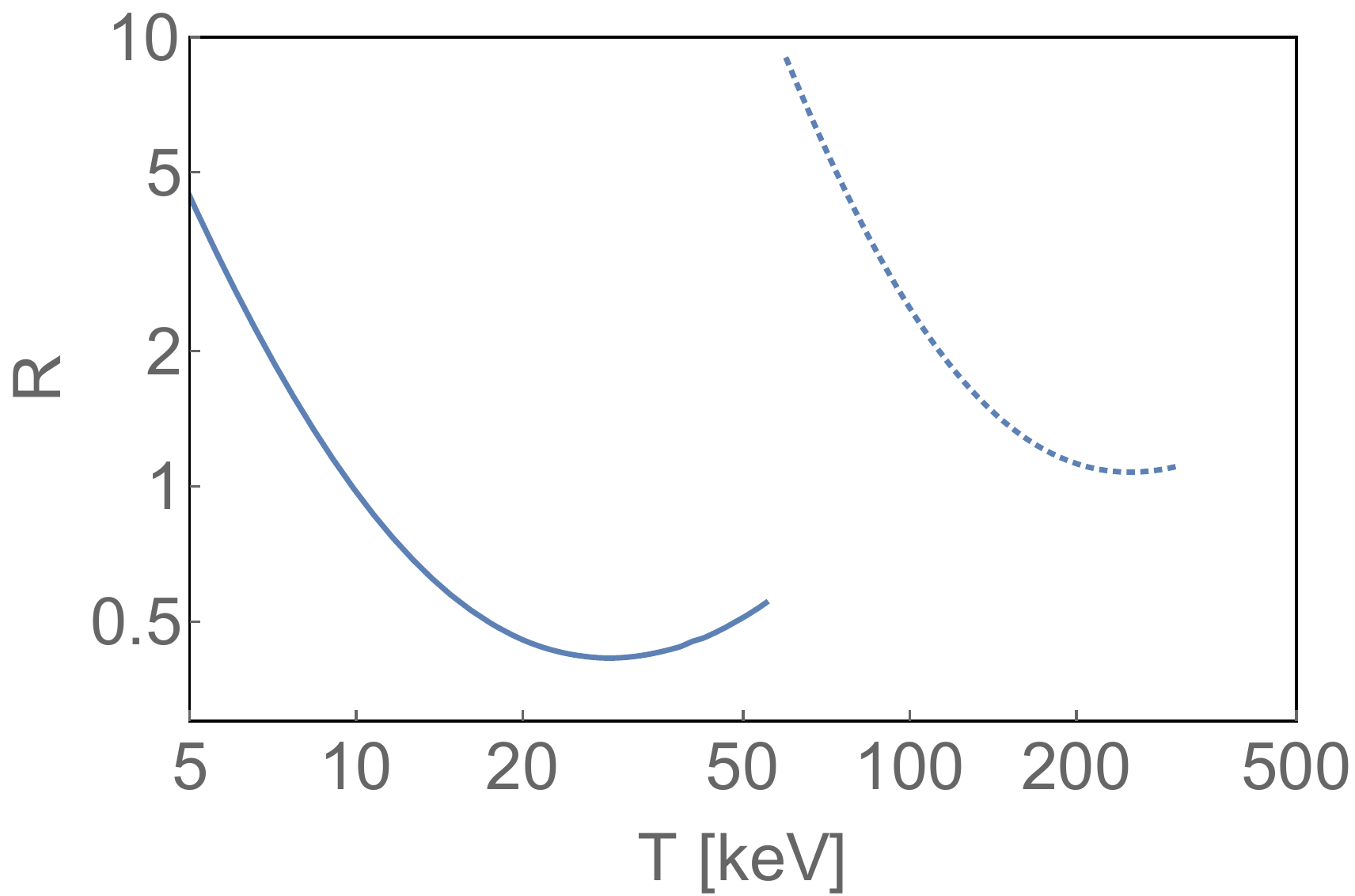}
\caption{$R$ metric plotted for constant-temperature DT (solid line) and \pb (dotted) plasmas. The plotting intervals are restricted by the validity of thermal reactivity fits and the incursion of the resonant beam into the bulk of the distribution function. We allow $\epsilon_{\rm beam}>2T$.}
\end{figure}

In contrast, the \pb cross section energy peak is close to the temperatures relevant to thermonuclear fusion. Likewise, due to the narrow resonance, substantial reactivity is gained by channeling fusion power to a resonant beam (e.g. the beam reactivity is 15.7X greater than thermal at $T=100\,{\rm keV}$). Typical values of $R$ in an igniting \pb plasma are slightly larger than unity, suggesting that a beam is a potentially useful investment in these systems.

The equilibration model employed here alters these conclusions somewhat. At low temperatures, the pressure is reduced and the beam reactivity benefits from the reduced thermal broadening of the fusion resonance. $R$ consequently improves, but the collisionality of the plasma increases simultaneously. Faster relaxation rates require the diversion of additional fusion power to maintain the beam and reduce the possible temperature separation between species, increasing the effective bremsstrahlung emission. Both of these effects limit the economic viability of a fast beam. However, the model excludes beam particles from the calculation of collision rates, reducing the collisionality of the plasma as the beam fraction is further increased. These competing collisional effects constitute the primary interactions of the beam with the thermal plasma.

Apart from the beam fraction, the other salient parameter in scanning plasma conditions is the number ratio $\epsilon$, defined in all cases as the ratio of heavy ions to light ions, including beam ions. An interesting divergence arose between the two reactions: in DT, $\epsilon\gtrsim1$ plasmas had lower ignition thresholds in both MCF and ICF. In \pb, $\epsilon<1$ plasmas were favored for MCF and $\epsilon>1$ plasmas were favored for ICF. In DT, the number of electrons does not depend on $\epsilon$, so the bremsstrahlung emission is mostly independent of this parameter. The slight preference for tritium-rich plasmas is likely seen in extremal cases because the same center of mass energy is available at lower pressures in a deuterium-rich plasma. In \pb, the preference for boron-poor plasmas in MCF is explained by the balancing of bremsstrahlung emission (increasing with $\epsilon$) and thermal reactivity (decreasing with $\epsilon$). 

The large boron density in the ICF optimum is harder to explain. The number ratio which maximizes the reactivity of a pressure-limited, constant-temperature system is $\epsilon=n_A/n_B=(1+Z_B)/(1+Z_A)=1/3$, in the case of \pbend. However, if the species temperatures are allowed to float, a high electron density (due to the boron excess) could suppress the electron temperature enough via bremsstrahlung emission that the reduction in the thermal content of the plasma outpaces the loss of reactivity and improves the volume gain in the aggregate ICF accounting.

In the presence of a beam, the $\epsilon$ and $T_i$ maximizing the total reactivity-pressure ratio should increase and decrease, respectively, as these changes create a rich environment of cold, resonant target ions for the lighter beam ions. The result is increased reactivity at fixed pressure. In \pb, the price of increasing $\epsilon$ is reduced thermal reactivity and, moreover, greater bremsstrahlung emission. In a constant-temperature model, both fuels behave as predicted; as the beam fraction rises from 0 to 1, there is a critical point where the beam reactivity exceeds the total thermal reactivity and the optimum operating point is an excess of very cold target ions.

Although the volume gain calculation is intrinsically conservative (cf. ignition in a hot spot), the assembly energies $E_a$ have been systematically underestimated, thereby exaggerating the gains reported here. In practice, it will be necessary to prepare the target at the prescribed temperature, which will likely involve heating from low temperatures (perhaps cryogenic, ambient at best). This heating will have to overcome bremsstrahlung and conduction losses; a better estimate of the volume gain will require tracing the target's evolution in $\rho R-T$ phase space as Lindl has done for DT hot spots in NIF targets.\cite{lindl95}

Investing substantial recirculated power in a fast beam has the potential to reduce ignition threshold conditions and ameliorate various engineering difficulties. The encouraging results of the stability analysis suggest that because the bulk ions are usually sufficiently cold, the necessary recirculated power can be regarded as small. In the case of DT plasmas, fast beams can increase the reactivity of a magnetically confined plasma at constant pressure or lower the $\rho R$ required for ignition by an inertially confined plasma. In \pb plasmas, the extra electron density increases the cost of maintaining a beam, limiting the gains which are possible in principle.

Indeed, this work found superior gains in DT with the introduction of a beam. In both MCF and ICF DT plasmas, these gains were substantial. A fast deuterium beam improved the DT reactivity by an order of magnitude while reducing the ion temperature to order-1 keV levels. However, the $10^4$ reduction in the ICF igniting assembly energy is arguably even more impressive. In \pb ICF plasmas, the corresponding reduction was only a factor of 30. However, this reduction is more impressive than the gains realized in \pb MCF, where a beam improved the total reactivity by only $20\%$. However, the alpha channeling which could be used to power the beam is likely critical to any steady state, magnetically confined scheme as a reliable means of abating the pressure poisoning effect. 

A key caveat to all the conclusions reached here, and the comparisons made between ICF and MFE as well as between DT and \pb fusion, is that these conclusions and comparisons are all based on an upper bound to a utility that in practice may be difficult to reach.  However, the utility that can possibly be reached serves as impetus to try to find ways to reach it.  The scenarios that we considered are in that respect at least not disallowed by the laws of physics, so they can serve at once not only as an impetus to achieve what is not disallowed, but also as a caution not to expect that more could be achieved.

Work  supported  by  DOE  Contract No.  DE-AC02-09CH11466 and DOE NNSA SSAA Grant No. DE274-FG52-08NA28553. M. J. H. was supported in part by the DOE NNSA SSGF under Grant No. DE-FC52-08NA28752.

\bibliography{mikebib2}

\appendix 

\section{Maintaining a monoenergetic beam}

There are several possible model choices for the recirculated power required to maintain the proton beam. Rider \cite{rider} calculated the amount of power necessary to prevent velocity space spreading of the beam past a prescribed thermal width $v_{th}\ll v_b$. This spread results primarily from beam-beam collisions; the Rider estimate of the recirculated power is proportional to $n_b^2$. This calculation is applicable when the beam lies far outside the thermal proton population and the velocity width of the beam is important.

In general, $P_{\rm recirc}$ must also counteract beam collisions on the thermal population, involving terms proportional to $n_b n_p$, $n_b n_B$, and $n_b n_e$. A more precise calculation would include both beam-beam and beam-thermal collisions and evaluate the reactivity integrals with a realistic slowing down distribution for the beam protons. In pursuing an upper bound of the gains realizable from alpha channeling, it has been assumed that the monoenergetic beam collides only with thermal particles.

\subsection{Collisional evolution of beam}
The Fokker-Planck equation governs the collisional evolution of an initially monoenergetic beam. Drag and velocity space diffusion induced by each of the thermal species contribute to the beam's eventual thermalization. We invoke a test particle analysis to argue for the validity of the model employed here. Consider, for example, the case of a 600 keV proton slowing in a plasma with $T_p=T_B=120\,{\rm keV}$ and $T_e=80\,{\rm keV}$. The number ratio is $n_B/n_p=0.20$ and $n_e=3.5\times10^{13}\,{\rm cm^{-3}}$. Although the beam slows primarily on thermal protons until an energy scattering time has passed, the fusion rate falls significantly from its original value on a collisionless time scale.

A similar DT case would have a 125 keV deuteron slowing in a plasma with $T_D=T_T=11\,{\rm keV}$ and $T_e=24\,{\rm keV}$. The number ratio is $n_T/n_D=1.25$ and $n_e=7.5\times10^{13}\,{\rm cm^{-3}}$. The picture is altered here because the ideal beam energy (maximizing the beam-thermal \emph{reactivity}\cite{mikkelsen}) is somewhat larger than the fusion resonance, such that most of the beam fusion events come after about one collision time. However, the fusion rate remains larger than the thermal ion collision times throughout the slowing down process.

In all cases, the beam remains well outside the thermal ion distributions, and most fusion events occur on a collisionless timescale. We do not anticipate that collisional spreading of the beam will modify the power balance laid out here.

\subsection{Collisionless instability}
Let us expand on comments made in the Introduction concerning the stability of the monoenergetic beams. In particular, we will present suggestive criteria for ion beams propagating in either cold or warm plasmas. A more detailed stability analysis for a collisional three-component plasma is beyond the scope of this work.

In a collisionless two-component plasma, the stability threshold for Langmuir waves in the presence of an ion beam is\cite{friedwong}
\begin{align}
\mathcal{D}(u)=Z'(u) +n \alpha^2 &Z'[\alpha(u-V)]\\ \nonumber
&-2(n+1)T=0,
\end{align}
where $Z'$ is the first derivative of the plasma dispersion function\cite{friedconte}, $u=\omega/kv_{th,i}$ is a dimensionless phase velocity, $\alpha=v_{th,i}/v_{th,b}$ describes the `coldness' of the beam, $V=V_b/v_{th,i}$ measures the beam's separation in velocity from the bulk ions, and $n=n_b/n_i$ is the dimensionless beam density, and $T=T_i/T_e$. Generally speaking, $V$ is $\mathcal{O}(1)$ in all configurations and $T$ is larger for \pb than DT plasmas (greater electron-ion temperature separation).

Taking asymptotes of Eq.~(A1), we can get some general notions of the stability of the beam-plasma configurations to be presented.\cite{friedwong} We will present the stability threshold as a curve in $(V,\alpha)$ space parametric in the phase velocity $u$.  For $V\gg1$, corresponding to a beam well outside of the thermal ion bulk (cold plasma), the stability threshold is
\begin{gather}
V=(\alpha+1)\sqrt{\frac{n}{2(n+1)T}},
\end{gather}
and likewise for $V\ll1$, a beam embedded in the thermal bulk (warm plasma),
\begin{gather}
V=\xi(1+n\alpha^3e^{-\alpha^2\xi^2}),
\end{gather}
with
\begin{gather}
\xi=\sqrt{\frac{n/2}{T(n+1)+1}}.
\end{gather}

In practice, these criteria may be regarded as a minimum amount of thermal spread intrinsic to the beam. Cold beams with little thermal spread (large $\alpha$) lying above these curves are unstable.

For example, we consider a case typical of the configurations to follow, a DT plasma with $n=0.1$ and $T=0.5$. In both DT and \pb plasmas, $V\gtrsim1$, so we will proceed cautiously using the $V\gg1$ asymptote. Then a stable beam has
\begin{gather}
\alpha<\sqrt{11}V-1.
\end{gather} 
For a representative $V=4$, $\alpha<4\sqrt{11}-1\approx12$, corresponding to a minimum beam temperature for stability of 100 eV in DT. (Typical beam energies are in the neighborhood of 127 keV, the center of mass energy at the fusion cross section peak.) Compare a typical case in \pb, with $n=0.1$ and $T=2$. We assume the beam-plasma interaction largely ignores the boron ions. The corresponding stability criterion is $\alpha<\sqrt{110}V-1\approx20$ for a typical $V=2$. The minimum thermal spread is then 1 keV, compared to a beam energy of about 627 keV.

In these representative examples, quasi-monoenergetic beams are well within the stability threshold. However, as the beam fraction is increased, the necessary stabilizing thermal spread may approach the beam energy in some scenarios.

\section{Optimized reactivity subject to maximum pressure}
In pressure-limited systems, the greatest achievable reactivity is an important experimental parameter that establishes confinement criteria. In the absence of a fast beam, the maximum of the function 
\begin{gather}
n_A n_B \left<\sigma v\right>\!(T)\,|\,n_A T+n_B T+n_e T<p_0
\end{gather}
is of interest. $n_A$ and $n_B$ are the number densities of the reacting ion species, such that $n_e=Z_A n_A+Z_B n_B$. $p_0$ denotes the maximum allowed pressure and $n_A/n_B=\epsilon$. Equivalently, one can maximize the Lagrange function $L=n_A n_B \left<\sigma v\right>-\lambda(n_A T+n_B T-p_0)$. The components of the gradient are
\begin{gather}
\frac{\partial L}{\partial n_e}=(1+Z_B+\epsilon(1+Z_A))T\lambda+2\frac{n_e\epsilon\left<\sigma v\right>\!}{Z_B+Z_A\epsilon}\\
\frac{1}{n_e}\frac{\partial L}{\partial \epsilon}=(Z_B-Z_A)T\lambda+\frac{Z_B-Z_A\epsilon}{Z_B+Z_A\epsilon}n_e \left<\sigma v\right>\! \\
\frac{1}{n_e}\frac{\partial L}{\partial T}=(1+Z_B+\epsilon(1+Z_A))\lambda+\frac{n_e\epsilon\left<\sigma v\right>'\!(T)}{Z_B+Z_A\epsilon}\\
\frac{\partial L}{\partial \lambda}=\frac{1+Z_B+\epsilon(1+Z_A)}{Z_B+Z_A\epsilon}n_e T-p_0\\ \nonumber
\end{gather}
Considering the complementarity condition, $\lambda=0$ implies $n_e\epsilon\left<\sigma v\right>\!=0$, which cannot hold in a finite-temperature plasma. Proceed, fixing $p=p_0$. Then
\begin{gather}
\epsilon=\frac{1+Z_B}{1+Z_A}\\
n_e=\frac{(Z_A+Z_B+2Z_A Z_B)}{2(1+Z_A)(1+Z_B) }\frac{p_0}{T}\\
\left<\sigma v\right>'\!(T)=2\frac{\left<\sigma v\right>\!(T)}{T}
\end{gather}
where the final equation can be solved implicitly for the $T=T_0$ which maximizes the reactivity.

\section{Evaluation of burn fraction $\phi$}
Without loss of generality, suppose initial number densities $n_{A0}$ and $n_{B0}$ such that $n_{A0}+n_{B0}=n_0$ and $n_{A0}/n_{B0}=\epsilon$. Then 
\begin{align}
n_{A0}=\frac{\epsilon}{1+\epsilon}n_0\\
n_{B0}=\frac{1}{1+\epsilon}n_0
\end{align}
and
\begin{align}
n_{A}(t)=\frac{\epsilon}{1+\epsilon}n_0-n\\
n_{B}(t)=\frac{1}{1+\epsilon}n_0-n
\end{align}
where $n=n(t)$ is the cumulative number of binary fusion reactions. Defining $\phi\doteq2n/n_0$,
\begin{gather}
\frac{dn}{dt}= \frac{n_0}{2}\frac{d\phi}{dt}=n_A n_B \left<\sigma v\right>\\  \nonumber
=\frac{n_0^2\left<\sigma v\right>}{4(1+\epsilon)^2}((\epsilon+1)\phi-2)(\epsilon(\phi-2)+\phi)
\end{gather}
which is a Riccati equation for $\phi$. Assuming constant $T_e$ and $T_i$ during burn, integrate over the confinement time, $R/3c_s$, to obtain the implicit equation
\begin{align}
\frac{1+\epsilon}{1-\epsilon}\log\left \lbrack \frac{\epsilon(\phi(\epsilon+1)-2)}{\epsilon(\phi-2)+\phi}\right \rbrack=\frac{R}{3c_s}n_0\left<\sigma v\right>
\end{align}
This can be inverted to obtain
\begin{align}
\phi=\frac{2}{\epsilon+1}\left[1-\frac{\epsilon-1}{\epsilon\exp\!\left( \frac{2}{H_B}\frac{\epsilon-1}{\epsilon+1}\rho R\right) -1}\right]
\end{align}
where $\rho=m_A n_{A0}+m_B n_{B0}$, $H_B=6 c_s m/\!\left<\sigma v\right>$, and $m=(\epsilon m_A+m_B)/(\epsilon+1)$.

\end{document}